\documentclass[%
 reprint,
 amsmath,amssymb,
 aps,
]{revtex4-2}

\usepackage{graphicx}
\usepackage{dcolumn}
\usepackage{bm}
\usepackage{xcolor} 
\usepackage{multirow}
\usepackage[colorlinks]{hyperref}
\hypersetup{%
	plainpages=true,
	breaklinks=true,
	hypertexnames=false,
	pageanchor=true,
	colorlinks=true,
	linkcolor={blue},
	citecolor={red},
	urlcolor={blue},
	anchorcolor={black}
}
\usepackage{siunitx} 
\usepackage[normalem]{ulem}

\begin{document}

\preprint{APS/123-QED}

\title{Deterministic generation of shaped single microwave photons \\ using a parametrically driven coupler} 

\author{Jiaying Yang$^{1,2}$}
\email{jiyang@chalmers.se}
\author{Axel Martin Eriksson$^{1}$}
\author{Mohammed Ali Aamir$^{1}$}
\author{Ingrid Strandberg$^{1}$}
\author{Claudia Castillo-Moreno$^{1}$}
\author{Daniel Perez Lozano$^{1}$}
\thanks{Present address: Imec, 3001 Leuven, Belgium}
\author{Per Persson$^{2}$}
\author{Simone Gasparinetti$^{1}$}
\email{simoneg@chalmers.se}
\homepage{https://202q-lab.se}

\address{$^1$Department of Microtechnology and Nanoscience, Chalmers University of Technology, SE-412 96, G\"{o}teborg, Sweden
\\$^2$  Ericsson research, Ericsson AB, SE-164 83, Stockholm, Sweden
}

\date{November 8, 2023}

\begin{abstract}
A distributed quantum computing system requires a quantum communication channel between spatially separated processing units. In superconducting circuits, such a channel can be realized by using propagating microwave photons to encode and transfer quantum information between an emitter and a receiver node. Here we experimentally demonstrate a superconducting circuit that deterministically transfers the state of a data qubit into a propagating microwave mode, with a process fidelity of \SI{94.5}{\%}. We use a time-varying parametric drive to shape the temporal profile of the propagating mode to be time-symmetric and with constant phase, so that reabsorption by the receiving processor can be implemented as a time-reversed version of the emission. We demonstrate a self-calibrating routine to correct for time-dependent shifts of the emitted frequencies due to the modulation of the parametric drive. Our work provides a reliable method to implement high-fidelity quantum state transfer and remote entanglement operations in a distributed quantum computing network.

\end{abstract}

\maketitle


\section{\label{sec:level1}\protect 
Introduction}

Superconducting circuits are one of the most favored approaches to realize a quantum computer~\cite{divincenzo2000physical, ladd2010quantum, nielsen2001quantum}. However, one challenge is to increase the scalability of the superconducting quantum circuits to outperform classical computers in solving computational problems. Aside from increasing the number of qubits on a single superconducting quantum circuit~\cite{arute2019quantum, jurcevic2021demonstration, gong2021quantum, wu2021strong}, a distributed quantum computing system~\cite{cirac1999distributed} or quantum internet~\cite{kimble2008quantum} that enables communication between remotely separated quantum processing units is also of great interest.
Many recent experiments have been devoted to realizing quantum channels between distributed quantum processors, based on phonons ~\cite{bienfait2019phonon, dumur2021quantum} channels, transducers between microwave and optical photons~\cite{bartholomew2020chip, wang2022generalized} and sorely microwave photons as discussed below.

Among the different realizations of quantum channels, one approach is to use propagating microwave photons as means to transfer quantum information between superconducting processors. In a single-rail encoding scheme, the microwave photon is emitted from one processor, transferred through the quantum channel, and then absorbed by the other processor.
In the emission process, quantum information is transferred from a stationary superconducting qubit of the emitter processor into a propagating microwave mode in a superposition state of zero and one photon~\cite{pechal2014microwave,  forn2017demand, kannan2022demand, li2023frequency}. When the propagating mode reaches the receiver processor, the time-reversed process is performed, leading to photon reabsorption and converting the propagating mode to a stationary qubit in the receiver~\cite{wenner2014catching}. Alternatively, the dual-rail encoding scheme has been explored as a method to detect photon loss in the channel. This approach has been experimentally implemented with heralding protocols using time-bin encoding \cite{kurpiers2019quantum, ilves2020demand, li2023frequency}. 

With the emission and reabsorption of the microwave photon, quantum state transfer and remote entanglement between the two quantum processors have been implemented deterministically~\cite{kurpiers2018deterministic, axline2018demand, campagne2018deterministic, leung2019deterministic, magnard2020microwave}. Adiabatic protocols~\cite{chang2020remote} have also been experimentally implemented to overcome channel loss. Building on these techniques, there are experimental realizations of quantum teleportation~\cite{steffen2013deterministic, chou2018deterministic, fedorov2021experimental}, photon gate sets~\cite{reuer2022realization}, and error detection protocols using multiphoton bosonic qubits~\cite{burkhart2021error}. In arguably the most advanced work in this area, a three-qubit Greenberger-Horne-Zeilinger state is deterministically transferred between two superconducting circuits by using a single quantum channel 3 times in a row~\cite{zhong2021deterministic}.

In this work, we demonstrate a superconducting quantum transceiver consisting of two transmon qubits and a flux-tunable parametric coupler, which deterministically emits single microwave photons into a coplanar waveguide \cite{besse2020realizing, kannan2022demand}. We shape the temporal profile of the emitted photon to be time symmetric, which enables the photon-reabsorption process at the receiver processor to be a reverse process of emission. For this purpose, we shape the amplitude of the emitted photon to be time symmetric and compensate for the temporal variation of its phase, caused by a time-dependent dispersive shift induced by the parametric drive. We perform quantum state tomography of the propagating microwave mode and quantum process tomography of the transfer operation, obtaining a process fidelity of \SI{94.5}{\%} for a time-symmetric mode envelope with an inverse hyperbolic cosine shape and a characteristic time $\tau =$ \SI{50.5}{ns}. Our device can be used as a building block in a distributed quantum computing system, operating either as an emitter or as a receiver.

\section{Methods}
\label{exp setup}
\begin{figure}\centering
\includegraphics[width=0.95\linewidth]{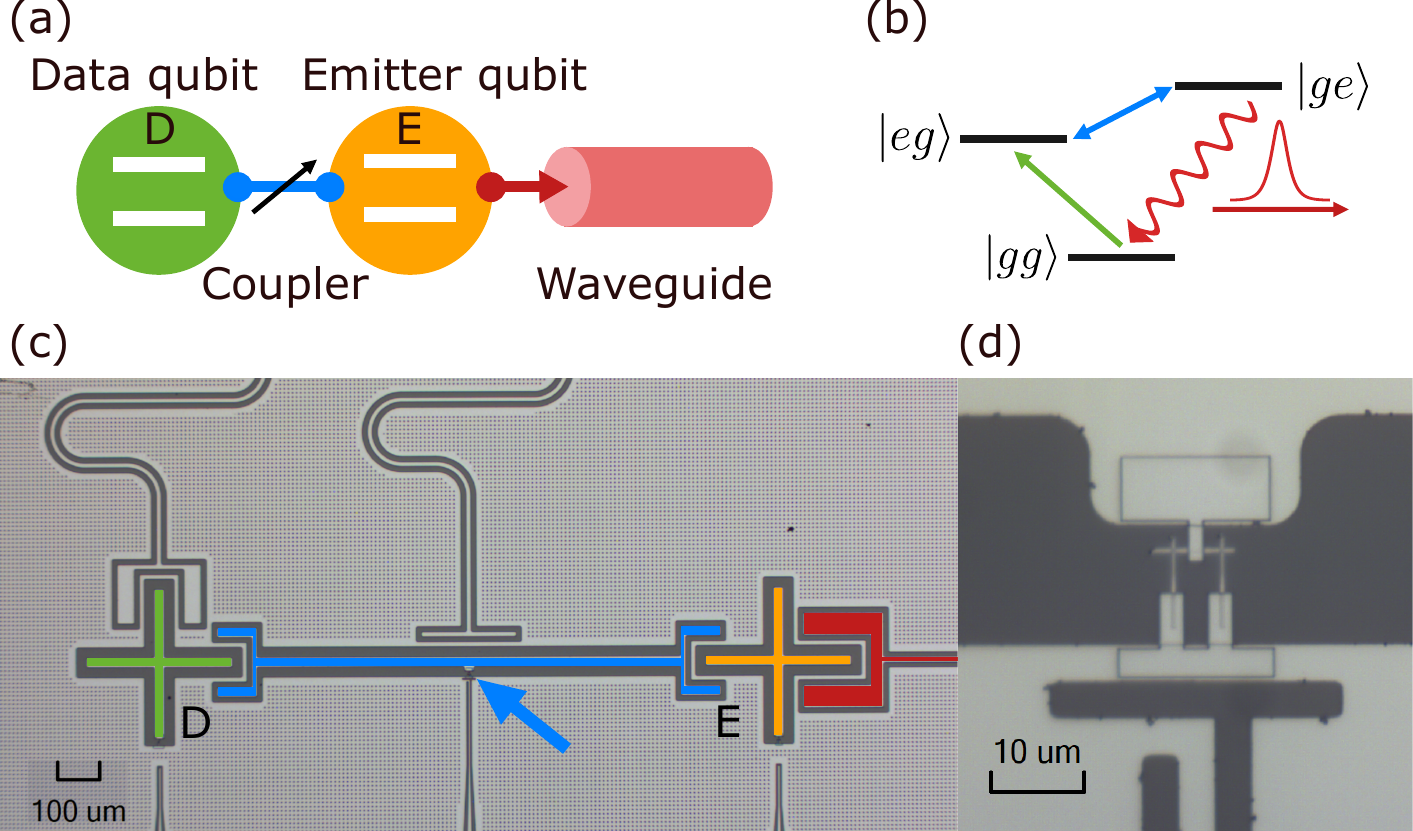}
\caption{Two-qubit quantum transceiver. (a) Schematic representation of the device. (b) Energy-level diagram illustrating the emission process. The state $|ij\rangle$ ($i,j=g,e$) refers to qubit D (E) prepared in state $i$ ($j$). (c) False-color optical micrograph of the device, which consists of a data qubit (green), an emitter qubit (orange), a flux-tunable coupler (blue), and a waveguide (red). (d) Enlarged view of the superconducting quantum interference device (SQUID) of the coupler, whose position is indicated by the blue arrow in (c).  }
\label{sample}
\end{figure}

Our device is a superconducting quantum circuit with two X-mon-type~\cite{barends2013coherent} transmon qubits~\cite{koch2007charge} --- a data qubit, D, and an emitter qubit, E --- capacitively coupled to a parametric coupler~\cite{mckay2016universal} between them [Fig.~\ref{sample}]. The design of the two-qubit system is similar to that presented in Ref.~\cite{bengtsson2020improved}.
Qubits D and E have transition frequencies $\omega_{\textrm{D}} /2\pi= 4.77$~GHz and  $\omega_{\textrm{E}} /2\pi= 4.95$~GHz, respectively. The coupler is frequency tuneable by an on-chip flux line, with a maximum frequency $\omega_{\textrm{C},0}/2\pi=7.735~\rm{GHz}$. Qubit E is capacitively coupled to a coplanar waveguide with a decay rate $\Gamma_{\rm E}/2\pi = 8$~MHz. Both qubit D and the coupler are capacitively coupled to a $\lambda/4$ coplanar waveguide resonator for characterization and readout measurements.  A summary of measured circuit parameters is presented in Table~\ref{parameters} of Appendix~\ref{setup}.

To characterize the transfer of quantum information from qubit D to a propagating mode, we first prepare qubit D in an arbitrary superposition state of the ground state $|g\rangle$ and the first excited state $|e\rangle$, cos$ \theta|g\rangle+e^{i\phi}$sin$ \theta |e\rangle$ [Fig.~\ref{sample}(b), green arrow], where the rotation angle $\theta$ is determined by the amplitude and length of the drive to qubit B, and $\phi$ is the phase of the drive.
Then we parametrically drive the coupler to induce an exchange interaction between qubits D and E 
[Fig.~\ref{sample}(b), blue arrow].
However, because qubit E is strongly coupled to the waveguide, it immediately decays into the ground state while emitting a microwave photon into the waveguide [Fig.~\ref{sample}(b), wavy red arrow]. As a result, the state of qubit D is transferred to a propagating mode of the waveguide, while qubit D is left in its ground state. We characterize the propagating mode by temporally matching the profile of the emitted photon, and perform quantum state tomography and quantum process tomography to reconstruct the density matrix of the mode and the Pauli matrix of the transfer process, respectively.

\section{Results}
\subsection{Parametric coupling}
\label{Para coupl}
\begin{figure}\centering
\includegraphics[width=0.95\linewidth]{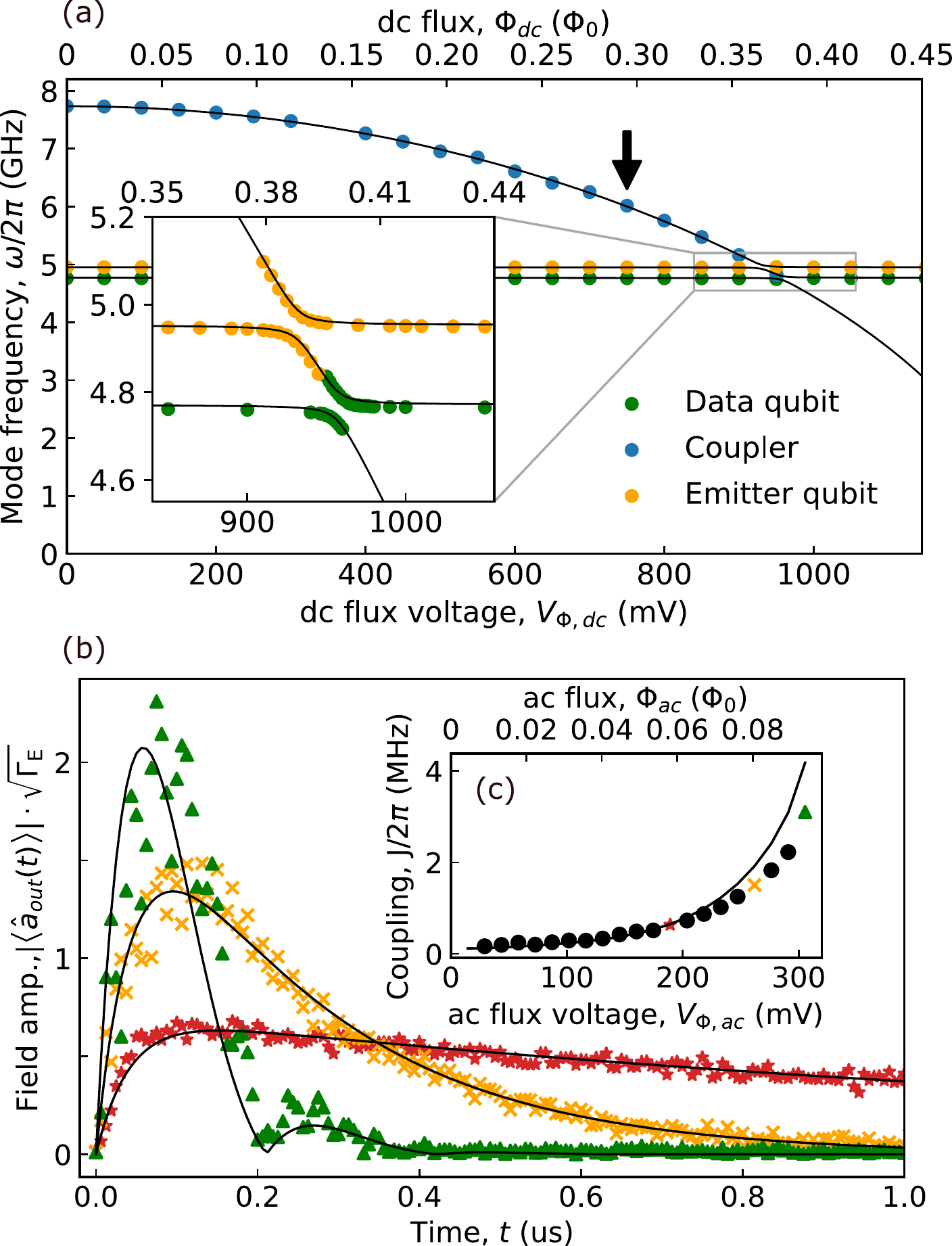}
\caption{Parametric driving of the flux-tunable coupler.  (a) Mode frequencies of the qubits and the coupler versus applied dc flux, $\Phi_{\textrm{dc}}$. The filled circles are the measured frequencies, and the solid curves are the theoretical prediction of the hybridized modes, based on the eigenvalues of the Hamiltonian in Appendix \ref{H eff eigenstates}. The black arrow indicates the operating point of the coupler,  $\Phi_{\textrm{dc}}=0.295\Phi_0$. The inset shows a zoom-in of the avoided crossing between the coupler and qubits.  
(b) Instantaneous amplitude of the emitted photon field versus time for constant, selected amplitudes of the coupler ac drive: $\Phi_{\textrm{ac}}= 0.057\Phi_{0}$
(red stars), $\Phi_{\textrm{ac}}= 0.079\Phi_{0}$
(yellow crosses), and $\Phi_{\textrm{ac}}= 0.092\Phi_{0}$
(green triangles). Solid lines are fits of the theory model.
(c) Qubit-qubit exchange interaction strength, $J$, as a function of the flux ac modulation amplitude, as extracted from the fits to the datasets in (b) (colored symbols) and from similar datasets taken at other amplitudes (black filled circles).  The black solid line is the corresponding theory prediction, see Appendix~\ref{app_model} for details. }
\label{fig_coupler}
\end{figure}

After preparing the desired state in qubit D,
the parametric drive of the coupler enables the exchanging of excitations between the two qubits. The frequency of the coupler is flux tunable, and the relationship between the coupler frequency and the applied magnetic flux $\Phi$ is approximately given by~\cite{koch2007charge}

\begin{equation} \label{fit_coupler}
\omega_{\textrm{C}}(\Phi) = \omega_{\textrm{C,0}}\sqrt{\mid \textrm{cos}(\pi \Phi/\Phi_0)\mid}\, ,
\end{equation}
where  $\omega_{\textrm{C,0}}$ is the frequency of the coupler when there is no flux bias, and $\Phi_0$ is the flux quantum. To exchange the desired state from qubit D to qubit E, we apply dc and ac currents to the flux line, resulting in a time-dependent magnetic flux~\cite{mckay2016universal}
\begin{equation} \label{flux DC and AC}
\Phi(t) = \Phi_{\textrm{dc}} +A(t) \textrm{cos}(\omega_{\textrm{m}} t)\, ,
\end{equation}
where the constant flux bias $\Phi_{\textrm{dc}}$ and the time-dependent envelope $A(t)$ determine the strength of the coupling rate $J$ between the two qubits. We first apply only the dc flux $\Phi_{\textrm{dc}}$ and measure the frequency of the three modes as a function of it  [Fig.~\ref{fig_coupler}(a), filled circles]. 
We then fix an operating point for $\Phi_{\textrm{dc}}$ [Fig.~\ref{fig_coupler}(a), black arrow], which we empirically determine by balancing the trade-off between larger coupling strengths, which we obtain by bringing the coupler closer in frequency to the qubits, and limited hybridization between the three modes, which we achieve by distancing the coupler's frequency from that of the qubits. We verify in Appendix~\ref{Purcell dacay} that the Purcell decay from qubit D to the more rapidly decaying  qubit E is not limiting the lifetime of qubit E at the chosen value of $\Phi_{\rm dc}$.
To parametrically induce an exchange interaction between qubits D and E, we apply an ac flux tone of frequency $\omega_{\textrm{m}}\approx \omega_{\textrm{E}}-\omega_{\textrm{D}}$, approximately matching the detuning between the frequencies of the two qubits.

To characterize the parametric coupling, we first prepare qubit D in state $(|g\rangle+ |e\rangle)/\sqrt{2}$  by setting rotation angle $\theta$ of qubit B to be $\pi/2$ and phase $\phi$ to be 0. Then we apply an ac flux tone of constant amplitude, $A(t)=\Phi_{\rm ac}$, and measure the average amplitude of the output field in the waveguide versus time, $\langle \hat a_{\rm out}(t)\rangle$ [Fig.~\ref{fig_coupler}(b)]. 
For low and moderate drive amplitudes $\Phi_{\rm ac}$, the emitted photon has a fast rise at the beginning and then decays exponentially [Fig.~\ref{fig_coupler}(b), red stars and yellow crosses]. At higher values of $\Phi_{\rm ac}$, the emission profile develops a second peak [Fig.~\ref{fig_coupler}(b), green triangles], signaling a regime in which Rabi oscillation between the two qubits is faster than the emission to the waveguide. Theoretically, it can be seen that this regime is attained when $J>0.25\Gamma_{\rm E}$ (see Appendix \ref{app_model} for details).

By fitting a simple model to these photon field amplitude curves measured at different modulation amplitudes $\Phi_{\textrm{ac}}$, with the coupling rate between the two qubits, $J$, as a free parameter (see Appendix \ref{app_model}), we obtain $J$ as a function of $\Phi_{\textrm{ac}}$ [Fig.~\ref{fig_coupler}(c)].
Each data point in Fig.~\ref{fig_coupler}(c) is measured at the optimal modulation frequency $\omega_{\textrm{m}}^{\textrm{opt}}$ of the corresponding $\Phi_{\textrm{ac}}$, where  $\omega_{\textrm{m}}^{\textrm{opt}}$ is defined as the ac flux tone frequency that gives us a constant emitted photon field phase (see Fig.~\ref{omega_m_vs_g} in Appendix \ref{Coupler_frequency_VS_Am} for its dependence on $\Phi_{\textrm{ac}}$). Fig.~\ref{fig_coupler}(b) shows that the time scale of photon emission is controlled by the modulation amplitude $\Phi_{\textrm{ac}}$ via the coupling rate $J$. By varying the strength of the modulation amplitude in time, it is possible to control the instantaneous decay rate of qubit D, and as a result, shape the temporal envelope of the emitted photon.

\begin{figure}\centering
\includegraphics[width=0.95\linewidth]{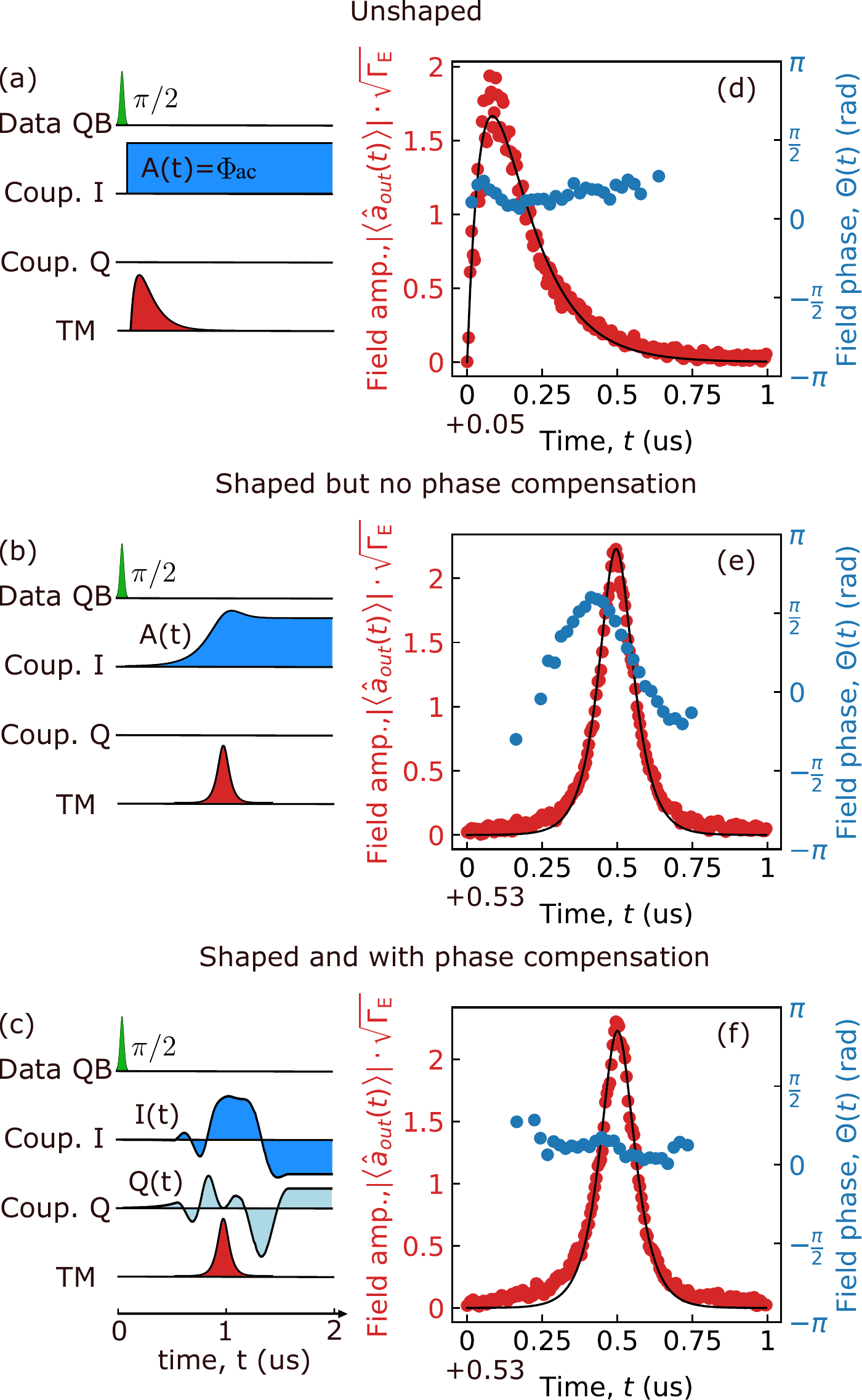}
\caption{Controlling temporal envelope and phase of the emitted photons. (a-c) The pulse sequence to measure the emitted photon under the three cases, (a) without reshaping, measured by using a constant coupler pulse with amplitude $0.079 \Phi_{0}$;
(b) with reshaping, measured by using a coupler pulse defined as Eq.~\eqref{custom pules} and amplitude $0.089 \Phi_{0}$, but no phase compensation and (c) with reshaping and phase compensation. (d-f) 
Normalized amplitude $\sqrt{\Gamma_{\rm E}}|\left<\hat a_{\textrm{out}}(t)\right>|$ and phase $\Theta(t) = \textrm{arg}[\left<\hat a_{\textrm{out}}(t)\right>]$ of the photon output field, measured under the three cases (a-c), respectively. $\Theta(t)$ is truncated such that the phase is not shown when the amplitude is close to zero ($\leqslant 0.05$). The time delay of photon emission relative to the completion of the $\pi/2$ pulse is represented by the +0 and +0.48 marked on the time axis of the (d-f) curves. }
\label{photon and pulse seuqnce}
\end{figure}

\subsection{Photon shaping and phase compensation}
\label{shaping}
To facilitate photon reabsorption at the receiver, it is desirable that the emitted photon has a time-symmetric envelope. In this case, a receiver identical to the emitter can reabsorb the photon with high efficiency when subjected to a time-reversed control sequence~\cite{cirac1997}.
We choose to shape the photon into the temporal profile with the expression $1/{\textrm{cosh}}(t/\tau)$, which enables the fastest falloff of the emitted photon envelope, by controlling the coupling rate $J$ between the qubits. To achieve this, the optimal envelope $A(t)$ of the ac flux tone sent to the coupler can be derived analytically~\cite{pechal2016microwave} to be

\begin{equation}
\label{custom pules}
    A(t) = \frac{\Gamma_{\textrm{eff}}}{4\textrm{cosh}(\Gamma_{\textrm{eff}}t/2)}\frac{1-e^{\Gamma_{\textrm{\textrm{eff}}}t}+(1+e^{\Gamma_{\textrm{eff}}t})\Gamma_{\rm E}/\Gamma_{\textrm{eff}}}{\sqrt{(1+e^{\Gamma_{\textrm{eff}}t})\Gamma_{\rm E}/\Gamma_{\textrm{eff}}-e^{\Gamma_{\textrm{eff}}t}}}\, ,
\end{equation}
where $\Gamma_{\textrm{eff}}$ is the effective decay rate of qubit D via qubit E, and $\Gamma_{\textrm{eff}}\leqslant \Gamma_{\rm E}$. 
Equation~\eqref{custom pules} is valid when the coupling rate $J$ is linearly proportional to the ac flux tone amplitude $\Phi_{\rm ac}$. However, in our experiment, we find that  we can observe highly symmetric photon emission using the pulse envelope given by Eq.~\eqref{custom pules}, despite our observed nonlinear dependence of $J$ on $\Phi_{\rm ac}$ [Fig.~\ref{fig_coupler}(c)], and we use this shape for simplicity. By using Eq.~\eqref{custom pules} to prepare $A(t)$, the photon envelope is shaped close to the desired symmetric shape $1/{\textrm{cosh}}(t/\tau)$ [Fig.~\ref{photon and pulse seuqnce}(e), red curve]. More accurate control over the functional form of the temporal envelope could potentially be achieved by considering the nonlinear dependence of $J$ on $\Phi_{\rm ac}$.

The envelope of the coupler flux pulse is prepared using $\Gamma_{\textrm{eff}}=0.25\Gamma_{\textrm{E}}$, which represents the fastest photon emission achievable in the device, as it is the largest $\Gamma_{\textrm{eff}}$ before observing a second peak due to a large coupling rate $J$ [Fig.~\ref{fig_coupler}(b), green triangles]. For this case, $A(t)$ is shown in the second row of Fig.~\ref{photon and pulse seuqnce}(b). Another benefit of this coupler pulse shape is that it has a sustained large value over time, which acts as a reset of qubit D. The pulse sequence for the shaped photon is shown in Fig.~\ref{photon and pulse seuqnce}(b) with the $A(t)$ set as Eq.~\eqref{custom pules} and its amplitude set to $0.089\Phi_{0}$, and the amplitude and phase of the shaped photon field are shown in Fig.~\ref{photon and pulse seuqnce}(e). 
After fitting the expression $1/{\textrm{cosh}}(t/\tau)$ to the measured field amplitude [Fig.~\ref{photon and pulse seuqnce}(e), solid curves], we get $\tau =$ \SI{50.5}{ns}. 
The comparison between the measured field amplitude to its theoretical expectation is shown in Appendix \ref{comparison photon measured and simulated}.

We calculate the symmetry $s$ of the 
photon field $\left<\hat a_{\textrm{out}}(t)\right>$ for both unshaped and shaped cases as~\cite{pechal2014microwave}
\begin{equation}
\label{cal symmetry}
    s = \underset{t_0}{\textrm{max}}\frac{\int |\left<\hat a_{\textrm{out}}(2t_0-t)\right>^*\left<\hat a_{\textrm{out}}(t)\right>|dt}{\int |\left<\hat a_{\textrm{out}}(t)\right>|^2dt}\, ,
\end{equation}
where $\left<\hat a_{\textrm{out}}(t)\right>$ contains both real and imaginary parts. The symmetry of the unshaped emitted photon is \SI{82.7}{\%}; however, it decreases to \SI{68.2}{\%} after shaping the photon (Table~\ref{table symmetry}). An inspection of the temporal profile of $\left<\hat a_{\textrm{out}}(t)\right>$ reveals that the decrease is mostly due not to an asymmetry in the amplitude of the signal, but rather to its varying phase [Fig.~\ref{photon and pulse seuqnce}(e), blue curve]. 
To make the photon reabsorption process a time-reversed process of photon emission, not only the amplitude profile of the emitted photon but also the phase need to be time symmetric, in contrast to what we observe in Fig.~\ref{photon and pulse seuqnce}(e). In the following paragraphs of this section, we demonstrate our capacity to control the phase of emitted photons by compensating the varying phase to be constant.

\begin{table}[h]
\caption[ht]{Symmetry of the emitted photon field $\left<\hat a_{\textrm{out}}(t)\right>$}
\label{table symmetry}
\centering
\begin{ruledtabular}
\begin{tabular}{c c}
                                    & s (\%) \\
\hline
Unshaped                            & 82.7 \\ 
Shaped but no phase compensation    & 68.2 \\ 
Shaped and with phase compensation  & 96.4 \\ 
\end{tabular}
\end{ruledtabular}
\end{table}

We understand the observed variation as follows. When we send an ac flux tone with a constant envelope to the coupler [Fig.~\ref{photon and pulse seuqnce}(a)], the time-averaged frequency of the coupler $\bar{\omega}_{\rm C}(\Phi_{\rm ac})$ decreases as the ac flux tone amplitude $\Phi_{\rm ac}$ increases. This is due to the nonlinear dependence between the coupler frequency and the flux applied to it [Fig.~\ref{fig_coupler}(a), blue curve]. Consequently, the frequencies of both qubits D and E are slightly pushed down due to their interaction with the coupler~\cite{ganzhorn2020benchmarking}. Hence, when applying the modulation pulse to the coupler, the frequency of the emitted photon will depend on the amplitude of the coupler drive $\omega_{\rm E}(\Phi_{\rm ac}) = \omega_{\rm D}(\Phi_{\rm ac}) + \omega_{\rm m}$. As long as the coupler amplitude $\Phi_{\rm ac}$ is constant, so is the frequency (or phase) of the emitted photon, as seen in Fig.~\ref{photon and pulse seuqnce}(d), blue curve. However, to shape the photon, we send a flux ac tone with a slowly varying time-dependent envelope $A(t)$ as in Eq.~\eqref{custom pules} [Fig.~\ref{photon and pulse seuqnce}(b), second row], which leads to a time-varying push on qubit D. As a consequence, the phase of the emitted photon has a corresponding time variation [Fig.~\ref{photon and pulse seuqnce}(e), blue curve], defined as $\Theta(t) = \textrm{arg} [\left<\hat a_{\textrm{out}}(t)\right>]$.

To compensate for the varying phase $\Theta(t)$ corresponding to a specific coupler ac tone envelope $A(t)$, we first characterize $\Theta(t)$ by measuring the emitted photon resulting from the drive with $A(t)$. Note that any constant shift of the average coupler frequency due to the profile $A(t)$ translates into a corresponding shift of the emitted frequency and can be trivially adjusted for. To compensate for the varying phase, the drive signal to the coupler is split into two quadratures on which we play the reversed varying phase  
\begin{align} \label{I_t Q_t}
I(t) &= A(t)\textrm{cos}\big(-2\pi\Theta(t)\big)\, ,\\
Q(t) &= A(t)\textrm{sin}\big(-2\pi\Theta(t)\big)\, .
\end{align}
The resulting drive can be seen in Fig.~\ref{photon and pulse seuqnce}(c). This protocol simplifies the calibration and operation process of the distributed communication system, 
 while still correcting for the varying phase to a high extent. 
After applying this phase compensation process, we obtain a constant phase for the emitted photon while keeping its envelope unchanged [Fig.~\ref{photon and pulse seuqnce}(f)]. The symmetry of the emitted photon calculated by Eq.~\eqref{cal symmetry} increases from \SI{68.2}{\%} to \SI{96.4}{\%} after compensating for the varying phase. 
Importantly, the phase variation is insensitive to which quantum state is emitted. Hence, once $\Theta(t)$ is characterized for a profile $A(t)$, it can be used to compensate for the varying phase for all quantum states to high accuracy.

\subsection{Tomography of the propagating mode}
\label{sec tomo}

\begin{figure}[ht!]
\centering
\includegraphics[width=0.95\linewidth]{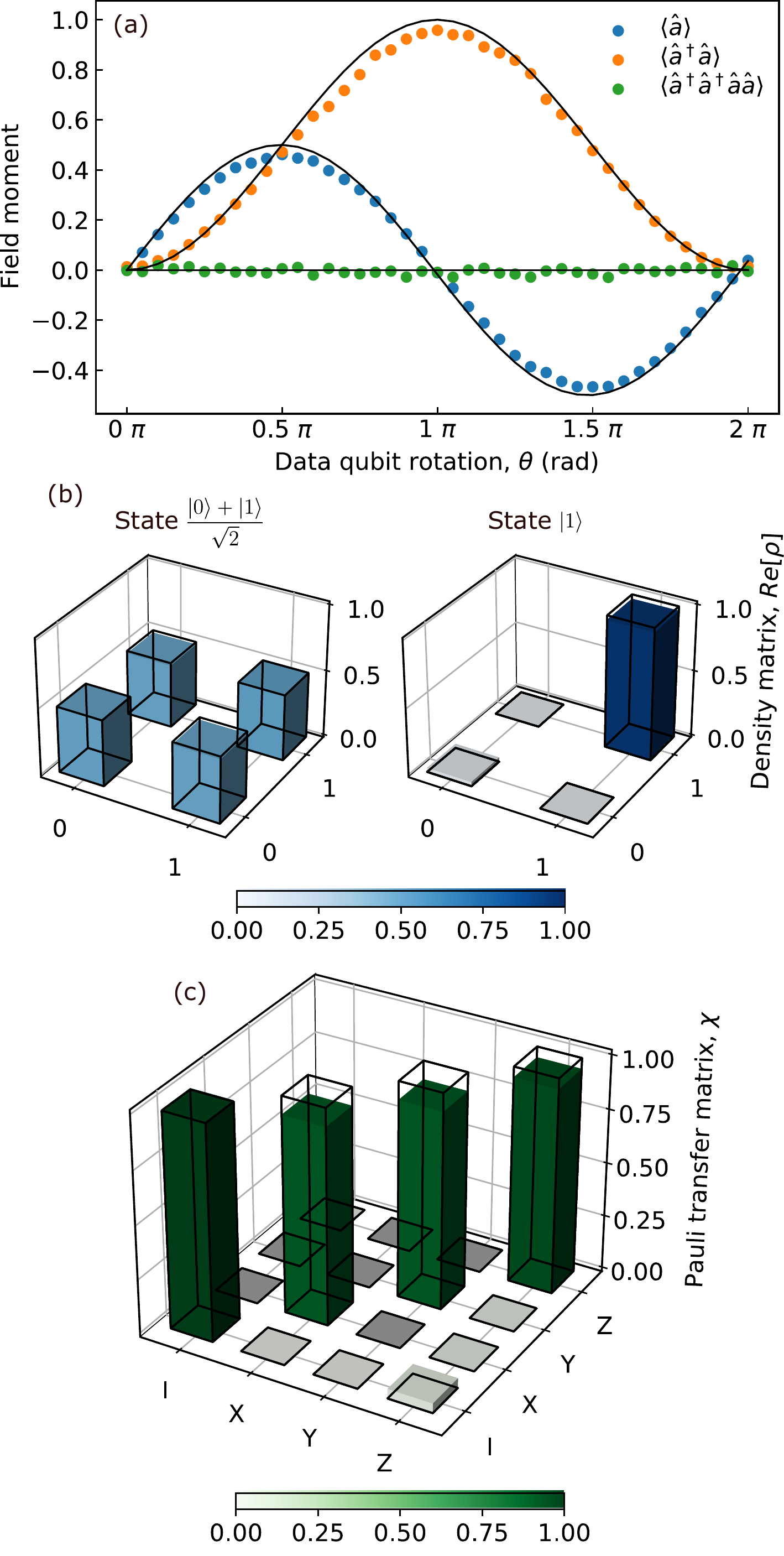}
\caption{Tomography measurement of the photon output field. (a) Selected moments of the emitted photon field. The filled circles are the measured data, and the black solid curves show the expected values of the moments for the ideal process. (b) Real part of the reconstructed density matrix for states $(|0\rangle+|1\rangle)/\sqrt{2}$ and $|1\rangle$, obtained from quantum state tomography. The blue bars show the measurement result and the black wireframe is the ideal case. The imaginary part of all matrix elements is smaller than 0.015. (c) Pauli transfer matrix for the data transfer between qubit D and the propagating mode, obtained from quantum process tomography measurement. The green bars show the measured result, and the black wireframe the identity process.}
\label{tomography}
\end{figure}

To characterize the propagating microwave mode, we perform quantum state tomography~\cite{steffen2006measurement}
using a linear amplification chain and temporal mode matching~\cite{eichler2011experimental}.
The amplification chain includes a traveling-wave parametric amplifier (TWPA)~\cite{macklin2015near} and a high-electron-mobility transistor (HEMT) amplifier, and we measured its quantum efficiency $\eta$ to be between $0.22$ and $0.3$ across the reported measurements, where $\eta = \frac{\frac{1}{2}}{\frac{1}{2}+n_{\rm added}}$ and $n_{\rm added}$ is the added noise photon number by the amplification chain. The template for mode matching $f(t)$ is chosen to have the same shape as the measured envelope of the emitted photon [Fig.~\ref{photon and pulse seuqnce}(a-c), bottom rows, labeled Template Matching (TM)]. 

We first calculate selected moments of the propagating mode $\hat a = \int \hat a_{\rm out}(t)f(t)dt$
as a function of the rotation angle of qubit D, $\theta$ 
[see Fig.~\ref{tomography}(a)], following Ref.~\cite{eichler2012characterizing}.
The moments are normalized according to the assumption that, when the parametric drive is turned on, the decay of qubit D occurs via two independent channels: the detected mode of the waveguide, and the intrinsic losses of qubit D, with rates $\Gamma_{\rm eff}$ and $\Gamma_{\rm D}$, respectively. Accordingly, when qubit D is prepared in state $|e\rangle$ in the unshaped case, we assume that $\langle \hat a^{\dagger} \hat a \rangle=\Gamma_{\rm eff}/(\Gamma_{\rm eff}+\Gamma_{\rm D})$. 
Based on this assumption, we calibrate the total gain of the amplification chain and normalize all data accordingly.
The mean amplitude of the field is represented by $\left<\hat a\right>$, which reaches the highest value at Rabi angle $\pi/2$. The mean photon number is represented by $\left<\hat a^{\dagger}\hat a\right>$, which takes its maximum when $\theta=\pi$. The fourth-order moment $\left<\hat a^{\dagger}\hat a^{\dagger}\hat a\hat a\right>$ is expected to be zero for all $\theta$ since a single photon is emitted. 
The normalized Glauber second-order correlation $g^{(2)}(0)=\left<\hat a^{\dagger}\hat a^{\dagger}\hat a\hat a\right>/(\hat a^{\dagger}\hat a)^{2}$ is \SI{0.024\pm0.027} and  \SI{-0.01\pm0.001} for states $(|0\rangle+|1\rangle)/\sqrt{2}$ and $|1\rangle$, respectively. Both values are very close to zero, indicating that the multiphoton component of the state is significantly smaller compared to the single-photon component. 

By initially preparing qubit D in states $(|g\rangle+|e\rangle)/\sqrt{2}$ and $|e\rangle$, exchanging the states to qubit E and performing quantum state tomography on the propagating mode of the emitted photon, we can reconstruct the density matrices for the two states, shown in Fig.~\ref{tomography}(b). We employ the convex optimization method from Ref.~\cite{strandberg2022simple} to reconstruct the density matrix that ensures that the resulting problem converges to the optimal solution. The fidelity of the density matrix is defined as~\cite{jozsa1994fidelity}
\begin{equation}
\label{fidelity definition}
    F(\rho, \rho') = \left(\mathrm{Tr}\sqrt{\sqrt{\rho}\rho'\sqrt{\rho}}\right)^2
\end{equation}

with $\rho$ being the target density matrix and $\rho'$ being the measured density matrix. 
The obtained fidelities of the density matrices are \SI{98.14\pm0.11}{\%} and \SI{96.80\pm0.36}{\%}, for the two states $(|0\rangle+|1\rangle)/\sqrt{2}$ and $|1\rangle$, respectively, with uncertainty computed from the standard deviation of repeated measurements. 
Further information on the quantum state tomography process is available in Appendix \ref{app tomography}.

We perform quantum process tomography~\cite{neeley2008process}  by preparing qubit D in the six cardinal states, $|g\rangle$, $|e\rangle$, $(|g\rangle+|e\rangle)/\sqrt{2}$, $(|g\rangle-|e\rangle)/\sqrt{2}$, $(|g\rangle+i|e\rangle)/\sqrt{2}$, and $(|g\rangle-i|e\rangle)/\sqrt{2}$, exchanging the state to qubit E and measuring the density matrices of the propagating modes with quantum state tomography. 
The optimized Pauli transfer matrix, derived from the six density matrices of the cardinal states, is depicted in Fig.~\ref{tomography}(c). 
The process fidelity is found to be \SI{94.534 \pm 0.002}{\%}, obtained by utilizing the Choi-Jamiołkowski correspondence between quantum operations and states \cite{jiang2013channel}. We define the process fidelity as the state fidelity ~\ref{fidelity definition} between the Choi matrices of the ideal and reconstructed processes.
This fidelity value is obtained by comparing it to the ideal process, the identity matrix.
In our experiment, the infidelity of both the photon state and the process mainly comes from the off-diagonal coherent terms of the reconstructed density matrices, limited by the longitudinal and transverse relaxation time of qubit D. As an example, when measuring state $(|g\rangle-|e\rangle)/\sqrt{2}$, we measured the off-diagonal terms of the reconstructed density matrix to be $0.45$, differing from the ideal value of $0.5$. This difference led to a lower fidelity for this state, \SI{94.85}{\%}, compared to the other cardinal states. A list of the fidelities for all six cardinal states is provided in Table \ref{table cardinal fidelities} in Appendix \ref{app tomography}.

\begin{table}[h]
\caption[ht]{Fidelity of the reconstructed density matrices from quantum state tomography}
\label{table fidelities}
\centering
\begin{ruledtabular}
\begin{tabular}{c c c}
                                    & $F_{(|0\rangle+|1\rangle)/\sqrt{2}}$ (\%)& $F_{|1\rangle}$ (\%)\\
\hline
Unshaped                     & 98.73(3)                       & 98.97(6)\\

Shaped but no phase comp.    & 94.73(28)                       & 94.77(78) \\

Shaped and with phase comp.  & 98.14(11)                     & 96.80(36)\\

\end{tabular}
\end{ruledtabular}
\end{table}

We also perform quantum state tomography for the other two cases, photon unshaped and photon shaped but without phase compensation (see Fig.~\ref{tomography 2} in Appendix \ref{app tomography} for the reconstructed density matrices, and Table 
\ref{table fidelities} for the fidelity of all three cases). The fidelity of the unshaped emitted photon is \SI{98.73\pm0.03}{\%} and \SI{98.97\pm0.06}{\%} for the two states $(|0\rangle+|1\rangle)/\sqrt{2}$ and $|1\rangle$, respectively. After shaping the temporal profile of the photon to time symmetric, the fidelities reduce to \SI{94.73\pm0.28}{\%} and \SI{94.77\pm0.78}{\%} because of the varying phase. After compensating for the varying phase, fidelities of \SI{98.14\pm0.11}{\%} and \SI{96.80\pm0.36}{\%} are achieved. 

The fidelity of the shaped and phase-compensated photons is slightly lower than that of the unshaped photons, primarily due to the larger time delay of the shaped photons and the further decay of qubit D during this delay. As shown in Fig.~\ref{photon and pulse seuqnce}(a-c), the unshaped photon's coupler pulse has constant amplitude over time, allowing for the maximum emission speed right from the start. In contrast, the shaped photon has a longer time delay, \SI{0.48}{\mu s}, due to the gradual increase in the amplitude of the coupler pulse envelope, which modulates the coupling strength between the two qubits to achieve a time-symmetric shape. For shaped photons, this longer delay exposes the qubit to additional decay of the state, consequently influencing the fidelity. The delay can be reduced, and consequently the fidelity improved, by reducing the rise time of the coupler pulse via numerical optimization.

\section{Discussion}
\label{Conclu}

In summary, we experimentally demonstrate the deterministic encoding from a stationary superconducting qubit to a propagating mode of an emitted microwave photon to a coplanar waveguide, based on a first-order parametric transition between two qubits connected by a coupler. We shape the emitted microwave photon to make its temporal profile time symmetric. However, the reshaping pulse causes time-dependent dispersive shifts of the qubits, making the phase of the photon output field uneven and reducing the fidelity of the emission process with respect to the time-symmetric target. To solve this problem, we execute phase compensation by implementing the reshaping pulse in both the I and Q quadratures and encoding the information of phase, which increases the fidelity of the photon field state. This reshaping and phase-compensation protocol enables the reabsorption process at the receiver to be a time-reversed version of the emission. Hence, the same procedure can be applied to the receiver processor. To characterize the propagating mode of the emitted photon, we perform both quantum state tomography and quantum process tomography on the photon output field and obtain a fidelity of \SI{94.5}{\%} for the encoding process.
Our results thus establish the structure developed in this work as a building block for distributed quantum computing or quantum network systems operating in the microwave frequency range.

In comparison with the previous works that utilize a qutrit-cavity structure and a second-order transition to control the emission~\cite{kurpiers2018deterministic,  kurpiers2019quantum, pechal2014microwave, ilves2020demand}, the presented technique uses a first-order parametric process. It requires no strong pump and therefore eliminates ac Stark shifts and heating of the data qubit, resulting in a system with fewer calibration procedures. Our photon shaping and phase compensation techniques can be straightforwardly implemented in devices with a similar structure, such as those that were used to demonstrate multipartite entanglement generation~\cite{besse2020realizing} and directional photon emission~\cite{kannan2022demand}. They can also be extended to various types of superconducting qubits as well as other solid-state qubits, such as quantum dots~\cite{stockklauser2017}. 

For future work, there are three potential further improvements on the speed of the emission and the phase compensation technique: first, by considering the nonlinear relationship between the coupling strength $J$ and the ac flux amplitude $\Phi_{\rm ac}$; second, by shortening the rise time of the coupler pulse, thereby reducing the exposure to T1 relaxation; and finally, by including the first-order dynamic corrections to the photon phase compensation.

\begin{acknowledgments}
The authors thank Andreas Bengtsson, Anita Fadavi, Christopher Warren, Daryoush Shiri, Marina Kudra, Mikael Kervinen, Shahnawaz Ahmed, and Yong Lu for technical assistance and useful discussion, and Lars Jönsson for his help in making the sample holders. The device in this work was fabricated in Myfab, Chalmers, a micro- and nanofabrication laboratory. The TWPA used in this experiment was provided by IARPA and Lincoln Labs. This work was supported by Ericsson Research and the Wallenberg Centre for Quantum Technology (WACQT).
\end{acknowledgments}

\appendix

\section{Experimental setup and device details}
\label{setup}
\begin{figure}
\centering
\includegraphics[width=0.95\linewidth]{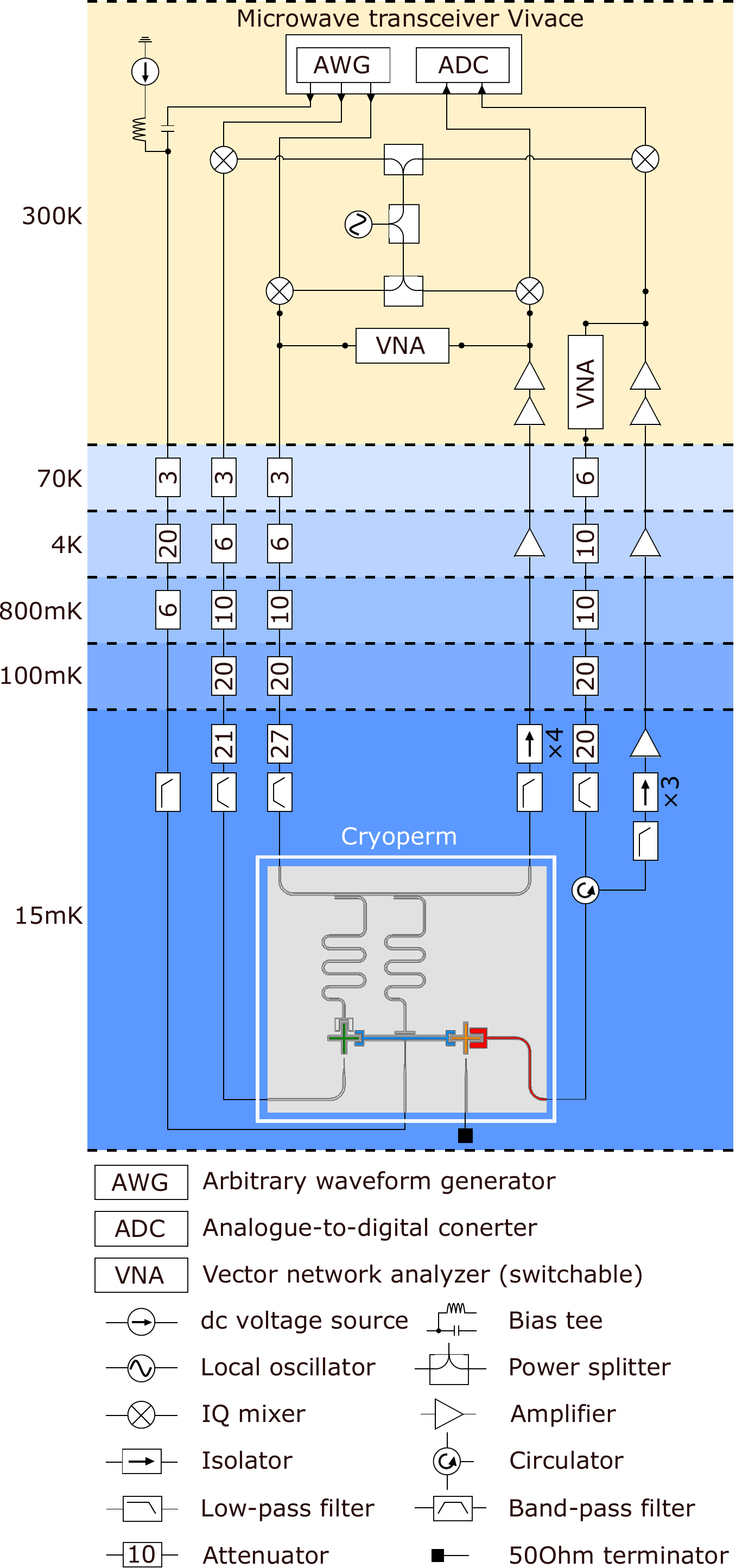}
\caption{Wiring diagram of the measurement setup. Note that the three outputs from the AWG and the two inputs from the ADC have two ports, including the in-phase component (I) and the out-of-phase component (Q), connected to the microwave transceiver board Vivace, but only one port is shown in the figure for simplicity. }
\label{wiring diagram}
\end{figure}

The $6.6$$ \times $$6.6\textrm{mm}^2$ device used in this work is fabricated on a silicon substrate, with the rf lines and the ground plane of the device made of aluminium deposited on top of the substrate. 
The device is wire bonded inside a copper sample holder, then put into a copper shield, and finally protected from the magnetic waves in a $\mu$-metal shield (cryoperm). The $\mu$-metal shield is mounted in the mixing chamber of a dilution refrigerator, to ensure all experiments are measured below the temperature $15$mK (Fig.~\ref{wiring diagram}).

The device is controlled with one transmission line, one charge line, one flux line, and one reflection input and output line at the waveguide. The two $\lambda/4$ resonators are coupled to the transmission line to read out the state of qubit D and the coupler. The charge line capacitively coupled to qubit D is used to prepare the desired state that will then be transferred. The flux line inductively coupled to the coupler SQUID loop enables frequency tunability. The waveguide capacitively coupled to qubit E connects to the reflection input and output lines on the other side, from where the emitted photon field is measured via a TWPA and a HEMT amplifier in the output line. Qubit E also capacitively couples to a charge line, which is not in use in this work. 

In the experiment, we operate both continuous-wave measurements and pulsed measurements. At room temperature, the vector network analyzer (VNA) is used for continuous-wave transmission and reflection measurements. The mode frequencies of qubit D and the coupler in Fig.~\ref{fig_coupler}(a) are obtained with two-tone spectroscopy measurements from VNA, and that for qubit E is obtained with single-tone spectroscopy with the same instrument. The power-dependent reflectance of qubit E is also measured with VNA, and by fitting the reflection we get the decay rate of qubit E, $\Gamma_{\rm E}$. The rest of the measured data, including the parametric drive of the coupler, the readout of the emitted photon output field, quantum state tomography and quantum process tomography measurements, are obtained by the pulsed measurements. We send microwave control pulses to gate the qubits with the arbitrary waveform generators (AWGs) and read out with the analogue-to-digital converters (ADCs) from the microwave transceiver
platform Vivace~\cite{Vivace}, after the up- (down-)conversion done by the IQ mixers and the local oscillators. 

The measured parameters of the device are shown in Table~\ref{parameters}, aside from which we also measure the coherence times of qubit D. When the coupler stays at the sweet spot, the longitudinal relaxation time $T_1$ and of qubit D is \SI{25.6}{\mu s}, with a standard deviation of \SI{10.4}{\mu s} measured over $12$ h; the transverse relaxation time $T_2$ and of qubit D is \SI{15.6}{\mu s}, with a standard deviation of \SI{4.7}{\mu s} measured over $12$ h. When the coupler is moved to the operating point by applying dc flux $\Phi_{\textrm{dc}}=0.295\Phi_{0}$
, those two $T_1'$ and $T_2'$ values are \SI{28.3}{\mu s} with variance \SI{9.2}{\mu s} and \SI{13.3}{\mu s} with variance \SI{3.7}{\mu s}. We have verified by simulation that the variance of the relaxation time has little impact on the time envelope of the emitted photon.

\begin{table}[h]
\caption[ht]{Measured device parameters. Here $f_R$ is the frequency of the readout resonator and $\kappa$ is the decay rate of the resonator to the feedline; $j$ is the coupling rate between qubit D (the coupler) and their readout resonator, while $g$ is the coupling rate between the qubits and the coupler; $\omega$ is the resonance frequency of the qubits and the coupler while no flux is applied to the flux line coupled to the coupler, and $\omega'$ is their frequencies at the operating point. The difference in transition frequency between the ground state and the first or second excited state, also known as the anharmonicity of the qubits or the coupler, is represented by $\alpha$. We denote by $\Gamma_{\rm E}$ the coupling rate between qubit E and the coplanar waveguide.}
\centering
\begin{ruledtabular}
\begin{tabular}{c c c c}
Parameter              & Data qubit & Coupler& Emitter qubit  \\
\hline
$f_R/2\pi$ (GHz)       & 6.172  & 6.72 & --          \\
$\kappa/2\pi$ (kHz)    & 632     & 772    & --          \\
$j/2\pi$ (MHz)         & 66.75   & 43.34  & --          \\
$\omega/2\pi$ (GHz)    & 4.771 & 7.735 & 4.953     \\
$\omega'/2\pi$ (GHz)   & 4.770   & 6.0    & 4.950        \\
$\alpha/2\pi$ (MHz)    & -224    & -62    &  \\
$g/2\pi$ (MHz)         & 33.85   & --     & 34.28      \\
$\Gamma_{\rm E}/2\pi$ (MHz)    & --      & --     & 8.0         \\
\end{tabular}
\end{ruledtabular}
\label{parameters}
\end{table}

\section{Model and simulation}
\label{app_model}

\subsection{Device Hamiltonian}
\label{app full hamiltonian}

The superconducting transceiver system can be modeled by the Hamiltonian of the two qubits and the coupler

\begin{equation} 
\label{H qubits and coupler}
\begin{split}
H(t)/\hbar = \omega_{\rm D}a_{\rm D}^\dagger a_{\rm D}+\omega_{\rm C}(t)a_{\rm C}^\dagger a_{\rm C}+\omega_{\rm E}a_{\rm E}^\dagger a_{\rm E}\\+g_{\rm D}(a_{\rm D}^\dagger a_{\rm C}+a_{\rm D} a_{\rm C}^\dagger)+g_{\rm E}(a_{\rm E}^\dagger a_{\rm C}+a_{\rm E} a_{\rm C}^\dagger)\\+\frac{1}{2}\alpha_{\rm D}(a_{\rm D}^\dagger)^2a_{\rm D}^2+\frac{1}{2}\alpha_{\rm C}(a_{\rm C}^\dagger)^2a_{\rm C}^2+\frac{1}{2}\alpha_{\rm E}(a_{\rm E}^\dagger)^2a_{\rm E}^2\, .
\end{split}
\end{equation}

The annihilation (creation) operators of the qubits and the coupler are represented by $a_{i}$ ($a_{i}^\dagger$) for $i = $ D, C, and E, whereas the anharmonicity of the qubits and the coupler are denoted by $\alpha_{i}$ for $i = $ D, C, and E. The coupling rate between qubit D (E) and the coupler is given by $g_{\rm D}$ ($g_{\rm E}$). The coupler frequency is time dependent since it is flux tunable as in Eq.~\eqref{fit_coupler} and we apply both the dc and ac flux as in Eq.~\eqref{flux DC and AC}.

Building on top of this Hamiltonian we solve the master equation by defining the collapse operator to be $\sqrt{\Gamma}\hat{\sigma}_{\rm E}^z$, with the \textsc{Python}
 package Qutip~\cite{johansson2012qutip}. In this master equation simulation, we use the initial density matrix of state $(|g\rangle+|e\rangle)/\sqrt{2}$ of qubit D and state $|g\rangle$ of qubit E as input and obtain the time evolution of the density matrix at each time point. We can obtain the expected shape of the emitted photon by monitoring the operator $a_{\rm E}$ of qubit E over time. Moreover, we can simulate how the photon shape changes for various modulation amplitudes $\Phi_{\rm ac}$.

\subsection{Obtaining mode frequencies using the effective Hamiltonian}
\label{H eff eigenstates}

In order to analyze the mode frequencies of the two qubits and the coupler as a function of the dc flux $\Phi_{\rm dc}$ [Fig.~\ref{fig_coupler}(a)], we begin by considering the effective Hamiltonian of the system

\begin{equation} 
\hat H_{\rm eff}/\hbar = 
\begin{pmatrix}
\omega_{\rm D} & 0 & g_{\rm D} \\
0 & \omega_{\rm E} & g_{\rm E} \\
g_{\rm D} & g_{\rm E} & \omega_{\rm C}(\Phi_{\rm dc})
\end{pmatrix}\, ,
\end{equation}

where $\omega_{\rm C}(\Phi_{\rm dc})$ is the frequency of the flux-dependent coupler, obtained by fitting the blue filled circles in Fig.~\ref{fig_coupler}(a) with Eq.~\eqref{fit_coupler}. By diagonalizing the Hamiltonian at various values of the dc flux $\Phi_{\rm dc}$, we obtain the three eigenstates of the matrix, which correspond to the frequencies of the three hybridized modes under the given $\Phi_{\rm dc}$, as shown by the solid curves in Fig.~\ref{fig_coupler}(a).

\subsection{Effective Hamiltonian of the qubits}

Upon considering the decay rate of qubit E, eliminating the coupler and restricting our analysis to the two qubits in the rotating frame, we obtain an effective non-Hermitian Hamiltonian~\cite{magnard2018fast, zhou2021rapid}

\begin{equation} 
\label{H qubits rf}
H_{\rm eff}/\hbar = 
\begin{pmatrix}
\omega & J \\
J & \omega-i\Gamma/2 
\end{pmatrix}\, ,
\end{equation}

where $J$ is the effective coupling strength between the two qubits and $\Gamma$ is the decay rate of qubit E. The decay rate of the amplitude, rather than the power, is taken into account in the element of qubit E by using $\Gamma/2$.

We start with preparing qubit D in state  $\alpha|g\rangle+\beta|e\rangle$ and qubit D in the ground state, which is 
\begin{equation} 
|\Psi(0)\rangle = \begin{pmatrix}
 \alpha\\
0
\end{pmatrix}\, .
\end{equation} 

After evolving time $t$, the state of qubit D becomes 
\begin{equation} 
|\Psi(t)\rangle = e^{-i\frac{\hat H_{\rm eff}}{\hbar}t}|\Psi(0)\rangle\, .
\end{equation} 

The expected output field $\left<\hat a_{\textrm{out}}(t)\right>$ of the emitted photon from qubit E can thus be calculated by

\begin{equation} 
\label{a out}
  \begin{aligned}
    \left<\hat a_{\textrm{out}}(t)\right> &= \sqrt{\Gamma} |\Psi(t)\rangle e^{i\omega t}\\
      & =-\frac{2i\alpha}{\sqrt{-16J^2+\Gamma^2}} \cdot \\
      &  
      \bigg(e^{-t(\Gamma-\sqrt{-16J^2+\Gamma^2})/4}-e^{-t(\Gamma+\sqrt{-16J^2+\Gamma^2})/4}\bigg)\, .\\
  \end{aligned}
\end{equation} 

 From this expression, we find that the photon output field is a subtraction of two exponential decays. One decays exponentially with rate $\frac{1}{4}(\Gamma+\sqrt{-16J^2+\Gamma^2})$, which dominates only in short time regime when $t$ is close to 0, and the other decays exponentially with rate $\frac{1}{4}(\Gamma-\sqrt{-16J^2+\Gamma^2})$ becomes dominant over extended time periods and effectively characterizes the decay rate of the photon field amplitude. This results in the amplitude of the photon field having a fast rise and then followed by an exponential decay when $J < \Gamma/4$ [Fig. \ref{fig_coupler}(b), red star and yellow cross curves].
 The decay rate that dominates at a longer time is the effective decay rate of the system, $\Gamma_{\rm eff} = \frac{1}{4}(\Gamma-\sqrt{-16J^2+\Gamma^2})$, and from where we deduce that the effective decay rate $\Gamma_{\rm eff} = 4J^2/\Gamma$ for $J\ll\Gamma$. The fastest effective decay rate $\Gamma_{\rm eff} = \Gamma/4$ is reached when $J = \Gamma/4$ and for larger $J$, we see a second peak in the emitted photon field because the oscillation between the two qubits becomes more dominant than the decay [Fig. \ref{fig_coupler}(b), green triangle curve]. 
 
 By obtaining the emitted photon amplitude under different flux ac amplitude $\Phi_{\rm ac}$ with Eq.~\eqref{H qubits and coupler} and the master equation simulation, fitting the amplitude with Eq.~\eqref{a out} and acquiring the coupling rate $J$ between the two qubits, we get the theoretical expectation of $J$ versus $\Phi_{\rm ac}$, depicted in Fig.~\ref{fig_coupler}(c) as the solid curve. The measured value of $J$ agrees with the theoretical expectation, as depicted in Fig.~\ref{fig_coupler}(c). Here, the red star, yellow cross, green triangle, and black filled circles represent data points obtained from the measured emitted photon amplitude. These points are derived by utilizing the same expression as the theoretical expectation, Eq.~\eqref{a out}, for the fitting.

 As cited from Ref.~\cite{mckay2016universal}, the effective Hamiltonian of the system is expressed as:
\begin{equation}
    H/\hbar= \frac{\Phi_{\rm ac}}{2}\frac{\partial J_{\rm dc}(\Phi)}{\partial\Phi}(\sigma_{\rm D}^X\sigma_{\rm E}^X+\sigma_{\rm D}^Y\sigma_{\rm E}^Y)
    \label{equation J_AC}
\end{equation}
where $\Phi_{\rm ac}$ is the amplitude of the ac flux pulse and  $J_{\rm dc}(\Phi)$ is the static coupling defined as:
\begin{equation}
J_{\rm dc}(\Phi)=\frac{g_{\rm D}g_{\rm E}}{2}(\frac{1}{\Delta_{\rm D}}+\frac{1}{\Delta_E})
\label{equation J_DC}
\end{equation}
with $\Delta_i=\omega_{\rm C}-\omega_i$, $i = \rm D, E$.

From Eq.~(\ref{equation J_AC}), we infer that the parametric coupling strength $J$ is proportional to $\Phi_{\rm ac}$.
For larger $\Phi_{\rm ac}$, the nonlinearity of the frequency-flux relation results in a nonlinear dependence of $J$ on $\Phi_{\rm ac}$, which we capture in Fig.~\ref{fig_coupler}(c) by full numerical simulations.


\section{Purcell decay estimation}
\label{Purcell dacay}

The Purcell decay $\kappa_p$ of qubit D into qubit E can be estimated as
\begin{equation}
    \kappa_p = (\frac{g_{\rm DE}+J_{\rm dc}}{\Delta_{DE}})^2\;\Gamma_{\rm E}
    \label{purcell}
\end{equation}
Here, $g_{\rm DE}$ is the capacitive coupling between the two qubits. It remains constant when we change the coupler frequency and can be approximated as
\begin{equation}
    g_{\rm DE}=\frac{C_{\rm DE}}{\sqrt{C_{\rm \Sigma, D}C_{\rm \Sigma, E}}}\sqrt{\omega_{\rm D}\omega_{\rm E}}
\end{equation}
where $C_{\rm DE}$ is the mutual capacitance of the two qubits, and $C_{\rm \Sigma, D}$($C_{\rm \Sigma, E}$) is the total capacitance of qubit D(E). We calculate $g_{\rm DE}/2\pi$=0.070MHz by knowing the capacitance matrix of the device from simulations. 

In Eq.~\ref{purcell},  $J_{\rm dc}$ represents the static coupling rate between qubits when dc flux is applied, and we calculate it according to Eq.~\ref{equation J_DC}. When the coupler is placed at 6GHz, $J_{\rm dc}/2\pi$ = 1.024MHz; $\Delta_{\rm DE}$ is the detuning between the two qubits, measured to be $\Delta_{\rm DE}/2\pi$ = 180MHz; $\Gamma_{\rm E}$ is the decay rate of qubit E into the waveguide, with a measured value of $\Gamma_{\rm E}/2\pi$=8MHz. Hence, we calculate the Purcell decay rate to be $\kappa_p/2\pi$ =0.296kHz, which is negligible compared to the measured decay rate of qubit D,  $\Gamma_{\rm D}/2\pi$=5.63kHz. Thus, we conclude that in the present configuration, the lifetime of qubit D is not limited by Purcell decay into qubit E.

\section{Two-dimensional sweep for the time-dependent coupler tone}
\label{Coupler_frequency_VS_Am}
In Sec. \ref{Para coupl}, we characterize the parametric coupling between the two qubits and in Fig.~\ref{fig_coupler}(c) we show the coupling rate $J$ as a function of the flux modulation amplitude $\Phi_\textrm{ac}$. To obtain $J$, we operate a two-dimensional sweep of both the flux modulation amplitude $\Phi_\textrm{ac}$ and the flux modulation frequency $\omega_\textrm{m}$ in the measurement of the averaged amplitude of the output field in the waveguide.

\begin{figure}[h]
\centering
\includegraphics[width=0.8\linewidth]{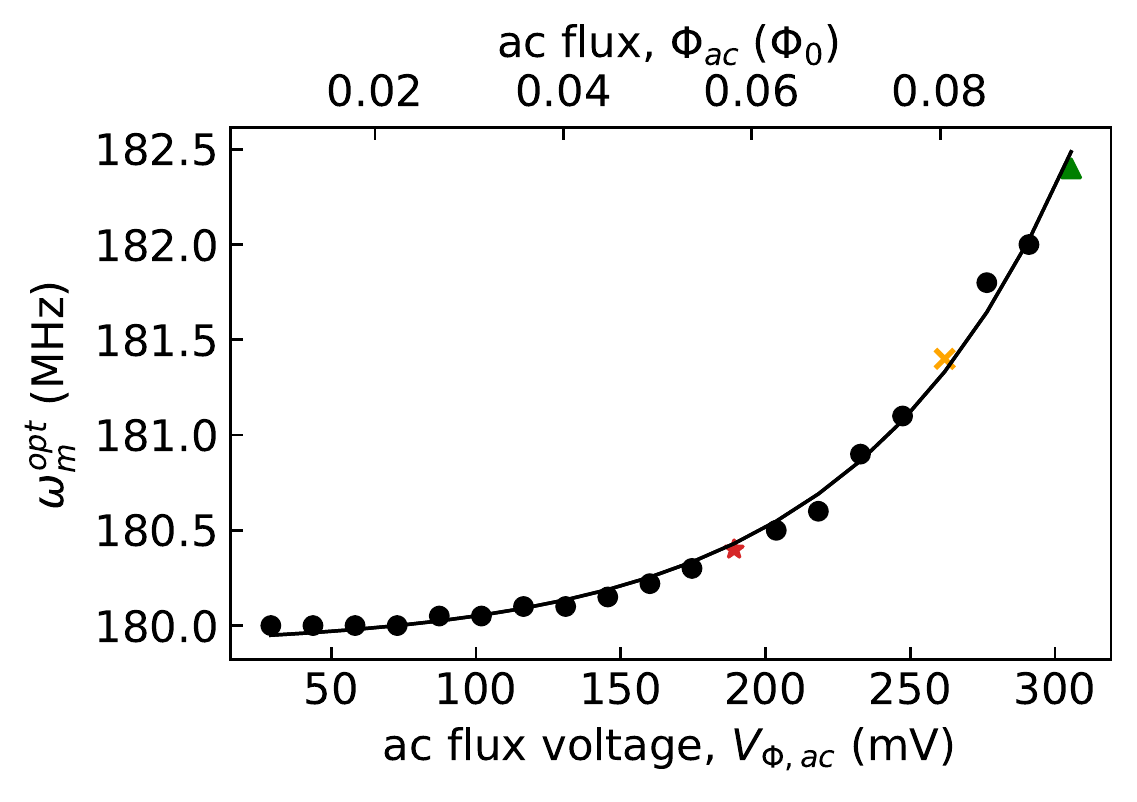}
\caption{The optimal modulation frequency $\omega_\textrm{m}^\textrm{opt}$ at each modulation amplitude. The red star, yellow cross, and green triangle correspond to the emitted photon field under the three cases in Fig.~\ref{fig_coupler} (b), and the black filled circles are similar datasets taken at other coupler ac drive amplitudes aside from the three.}
\label{omega_m_vs_g}
\end{figure}

To measure the output field of the emitted photon, the frequency of the ADC module is set at a 
 fixed value $\omega_\textrm{ADC}=\omega_\textrm{D}+\Delta_\textrm{DE}$, where $\Delta_\textrm{DE}=\omega_\textrm{E}-\omega_\textrm{D}$, a sum of qubit D pump frequency and the detuning between the frequencies of the two qubits, measured when only the dc flux tone is sent to the coupler. As discussed in Section \ref{shaping}, the frequencies of two qubits D and E shift slightly because of the parametric drive of the coupler and the dependence of the averaged coupler frequency $\bar{\omega}_{\rm C}(\Phi_{\rm ac})$ on $\Phi_\textrm{ac}$. At each modulation amplitude $\Phi_\textrm{ac}$, we fix $\omega_\textrm{D}$ and $\omega_\textrm{ADC}$ and sweep the flux modulation frequency $\omega_\textrm{m}$ around $\Delta_\textrm{DE}$ to measure the emitted photon. We are able to do this since qubit E has a large linewidth compared to 
 $\delta_\textrm{m}$, detuning between $\omega_\textrm{m}$ and $\Delta_\textrm{DE}$. The modulation frequency where we measure a flat phase is defined as the optimal modulation frequency $\omega_\textrm{m}^\textrm{opt}$ at a given $\Phi_\textrm{ac}$ (see Fig.~\ref{omega_m_vs_g} for $\omega_\textrm{m}^\textrm{opt}$ versus $\Phi_\textrm{ac}$), fitted by an exponential curve.

\section{Calibration of the ac flux amplitude}
When parametrically driving the coupler, aside from the dc flux tone, an ac flux tone is applied at the frequency of $\omega_\textrm{m}$. The total attenuation in the input and output line at this frequency is different from the attenuation at zero frequency when a dc flux tone is applied. Therefore, we calibrate the ac flux amplitude used in the measurement by multiplying it with a scaling factor to compensate for the attenuation difference.

To do the calibration, we first calculate the  averaged frequency of the coupler as a function of the ac flux amplitude $\Phi_{\rm ac}$ by taking the time integral of Eq.~\eqref{fit_coupler}

\begin{equation}
\label{omega_c vs amp}
\begin{split}
    \bar{\omega_{\textrm{C}}}(\Phi_{\rm ac}) &= \frac{1}{T}\int_{-T/2}^{T/2}\omega_{\textrm{C}}(t) dt \\
    &= \frac{1}{T}\int_{-T/2}^{T/2}\omega_{\textrm{C, 0}}\sqrt{\mid \textrm{cos}(\pi \Phi(t)/\Phi_0)\mid} dt\,,
\end{split}
\end{equation}

with $\Phi(t) = \Phi_{\rm ac}\textrm{cos}(\omega_\textrm{m}t)$ and $T = 2\pi/\omega_\textrm{m}$ one time period of the sinusoidal expression. After the time integral, we get the expression of the time-averaged coupler frequency as a function of the ac flux amplitude, independent of time and the flux frequency. 

 Then we investigate the relationship between the amplitude of an ac flux tone and the resulting coupler frequency in the experiment. As discussed above, the ac flux amplitude used in the experiment $\Phi_{\rm ac}'$ needs to be rescaled by factor $k$, so that the measured data can be fit by Eq.~\eqref{omega_c vs amp} with $\Phi_{\rm ac} = k\Phi_{\rm ac}'$. In our case, we obtain $k = 0.782$ from the fit. When transferring the flux amplitude from voltage $V_{\Phi, \textrm{ac}}$ (e.g., Fig.~\ref{omega_m_vs_g}, bottom x axis) to flux $\Phi_{\rm ac}$ (e.g., Fig.~\ref{omega_m_vs_g}, top x axis), we always apply the scaling factor $k$ by multiplying the voltage level with $k$ to compensate for the attenuation differences caused by various frequencies.

\section{Comparing the shaped photon in measurement and simulation}
\label{comparison photon measured and simulated}

The amplitude and phase of the shaped emitted photon are displayed in Figs.~\ref{photon and pulse seuqnce}(e)-\ref{photon and pulse seuqnce}(f), and we use the function $1/\textrm{cosh}(t/\tau)$ to fit its amplitude, which exhibits the symmetry of the photon envelope. In this section, we utilize the model introduced in Appendix~\ref{app full hamiltonian} to simulate the system and compare the measured photon field amplitude with the simulated theoretical expectation. 

\begin{figure}[h]
\centering
\includegraphics[width=0.8\linewidth]{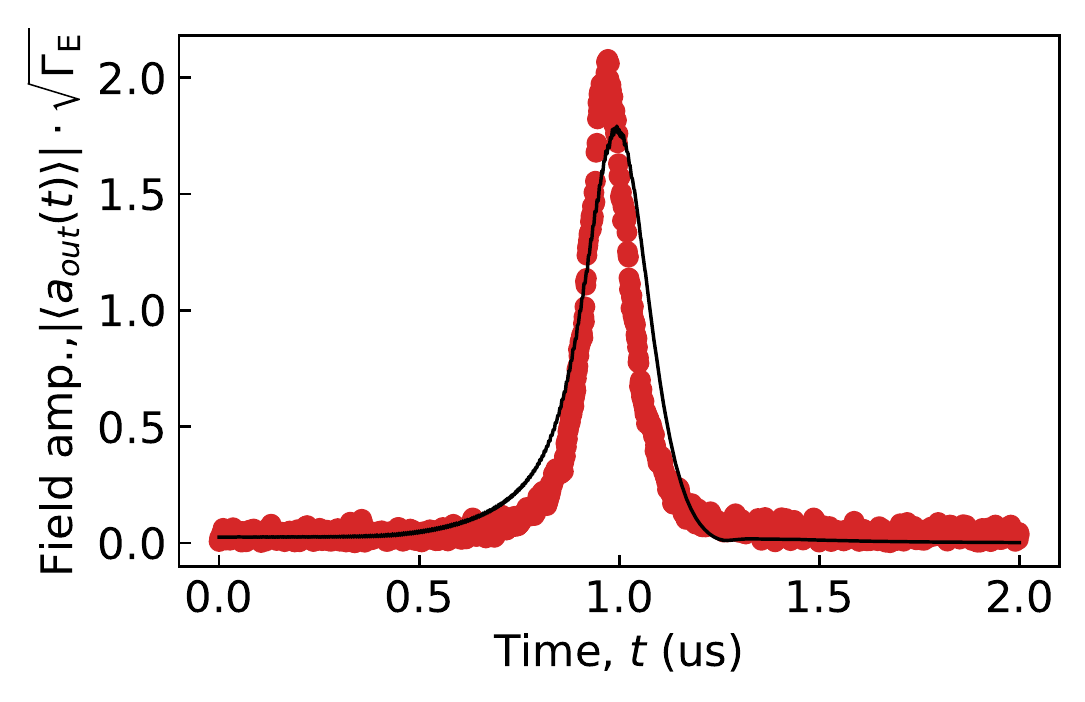}
\caption{The amplitude of the output field of the shaped emitted photon. The red filled circles are the measured data, and the solid curve is the theoretical simulation.}
\label{shaped photon comparison}
\end{figure}

Fig.~\ref{shaped photon comparison} displays the theoretical expectation of the photon field amplitude as a solid curve, as well as the measured data as red filled circles; the coupler pulse with an envelope defined by Eq.~\eqref{custom pules} is used for this comparison. The rising time, defined as the width of the photon envelope at \SI{10}{\%} of its highest point, is considered reasonable for the measured data to be \SI{77.5}{\%} of the simulation, taking into account the simplicity of the model.

\section{Readout and tomography}
\label{app tomography}
To perform quantum state tomography and quantum process tomography of the emitted photon, we apply a mode-matching relation to obtain the time-independent photon mode $\hat a$ from the measured time-dependent output field of the emitted photon $\left<\hat a_{\textrm{out}}(t)\right>$
\begin{equation} \label{a and a_t}
\hat a = \int\, dt\left<\hat a_{\textrm{out}}(t)\right>f(t)\,,
\end{equation}
with $f(t)$ the template for mode matching. The template $f(t)$ defines which mode from the continuum is probed in the tomography. Hence, the template should have the temporal profile of the photon field carrying the quantum information to match it with the highest efficiency. Thus, in the experiment, we prepare $f(t)$ to be a real template with the same shape as our emitted photon field amplitude [Fig.~\ref{photon and pulse seuqnce}(a-c), bottom row, labeled TM]. This choice is motivated by our objective of realizing a component that can be used as both emitter and receiver. To accomplish this, it is necessary to prove our capability to emit photons with symmetric amplitude and phase. However, when we apply a real, phase-fixed pulse to the coupler, we encounter a deterministic but unsymmetric phase variation. In Section III B, we show that we can gain control over the phase of the emitted photon by making the phase constant. This approach corresponds to a real template, which compares the emitted photon with the expected case for the emitter-receiver system. The template is also normalized to fulfil $\int dt|f(t)|^2=1$, to guarantee the bosonic commutation relation $\left [ \hat a, \hat a^{\dagger} \right ]=1$.

The photon mode $\hat a$ cannot be directly detected from measurements of the emitted photon field due to added noise from e.g.\ imperfect mode matching, cable losses, and amplification noise. Thus, by following the pulse sequences in Fig.~\ref{photon and pulse seuqnce}(a-c), we measure a complex mode $\hat{S}$ that includes both the photon mode $\hat a$ and an effective noise mode $\hat h^\dagger$:
\begin{equation}
    \hat{S}=\hat{X}+i\hat{P}:=\hat a+\hat h^{\dagger}\, .
\end{equation}
Operator $\hat{S}$ represents the signal of the two conjugating quadratures that we measured, $\hat{X}$ and $\hat{P}$. The moments $\langle (\hat{S}^\dagger)^n\hat{S}^m \rangle$ are a function of the moments of the photon mode and noise mode~\cite{eichler2011experimental}
\begin{equation}
\label{moments calculation}
    \left< (\hat{S}^\dagger)^n\hat{S}^m\right> = \sum _{i,j = 0}^{n, m}\binom{m}{j}\binom{n}{i}\left< (\hat a^\dagger)^i\hat a^j\right>\left< \hat h^{n-i}(\hat h^\dagger)^{m-j}\right>\, ,
\end{equation}
where $m$ and $n$ are integers $\geqslant 0$. 
We conduct measurements on the total mode in two situations, namely $\hat{S}_\textrm{on}$ and $\hat{S}_\textrm{off}$, via interleaved switching between the measurements with and without the coupler time-dependent tone. The $\hat{S}_\textrm{off}$ situation solely comprises the noise mode $\hat h^\dagger$. We measure both situations in the interleaved way for $5\times 10^6$ shots, and obtain a two-dimensional histogram from each case, with the real part of the total mode as the $x$ axis and the imaginary part of it as the $y$ axis. We assume that $\langle n \rangle = \hat a^{\dagger}\hat a=\Gamma_{\rm eff}/(\Gamma_{\rm eff}+\Gamma_{\rm D})$ when qubit D is prepared in state $|e\rangle$ in the unshaped case, and normalize the histograms for all three cases accordingly. By doing a Gaussian fit to the line slice of the normalized noise histogram, we get the noise photon number of the system. The moments of the emitted photon field are linearly calculated from the moments of $\hat{S}_\textrm{on}$ and $\hat{S}_\textrm{off}$ according to Eq.~\eqref{moments calculation}.

After preparing histograms from $\hat{S}_\textrm{on}$ and $\hat{S}_\textrm{off}$, the tomography problem can be considered as obtaining the density matrix $\rho$ from solving the linear equation $A\overrightarrow{\rho}=\overrightarrow{b}$, where $\overrightarrow{\rho}$ (size M$\times$1, where M is 4 in our case) is the flattered density matrix;  $\overrightarrow{b}$ is the flattened histogram $\hat{S}_\textrm{on}$ (size N$^2\times$1, where N is the number of bins of the histogram on both the real and imaginary axes) and $A$ (size M$\times$N$^2$) contains information of measurement settings. The matrix $A$ is obtained from the positive operator-valued measure (POVM) and the basis operators. With $A$ and $\overrightarrow{b}$ known, we feed it into convex optimization protocol to obtain the reconstructed density matrix of the state~\cite{strandberg2022simple}. 

\begin{figure}[h]
\centering
\includegraphics[width=0.9\linewidth]{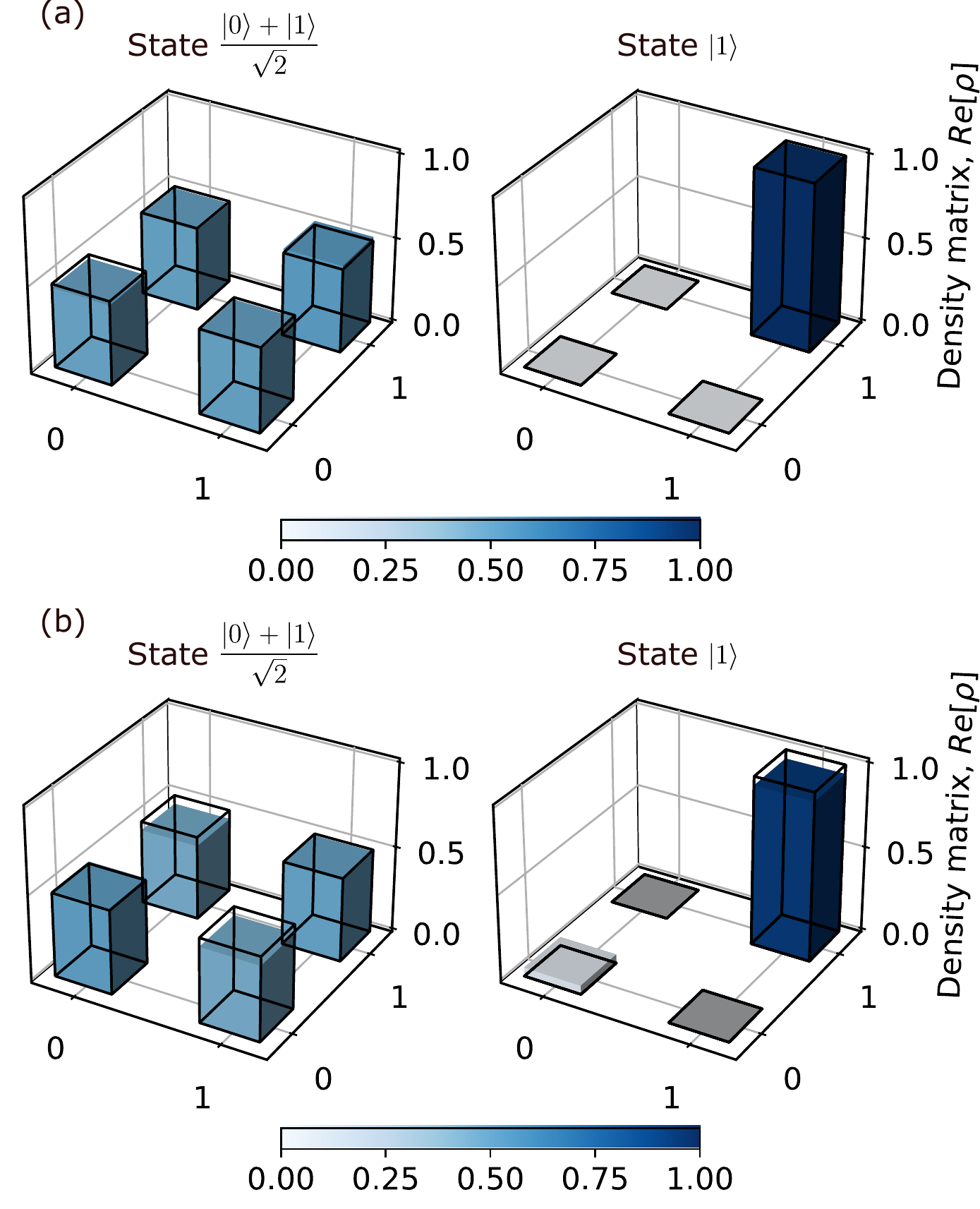}
\caption{Real part of the reconstructed density matrices for (a) unshaped and (b) shaped but no phase compensation cases, where the blue bars show the measured result, and the black frame shows the expected. For both cases, the imaginary part of all matrix elements is smaller than 0.015.}
\label{tomography 2}
\end{figure}

In Sec. \ref{sec tomo}, we compare the quantum state tomography for the emitted photon field measured in three cases: (i) without shaping, (ii) with shaping but without phase compensation, and (iii) with both shaping and phase compensation. The reconstructed density matrices of the first two cases are shown in Fig.~\ref{tomography 2}, and that of the last case is shown in Fig.~\ref{tomography}(b) in Sec. \ref{sec tomo}.

\begin{table}[h!]
\caption[ht]{The fidelity of the reconstructed density matrices corresponding to the six cardinal states. }
\label{table cardinal fidelities}
\centering
\begin{ruledtabular}
\begin{tabular}{l l }
$F_{|0\rangle}$                        & \SI{99.94\pm0.0001}{\%}  \\ 
$F_{|1\rangle}$                        & \SI{96.12\pm0.0065}{\%}  \\ 
$F_{(|0\rangle+|1\rangle)/\sqrt{2}}$   & \SI{96.18\pm0.0001}{\%}  \\ 
$F_{(|0\rangle-|1\rangle)/\sqrt{2}}$   & \SI{94.85\pm0.0168}{\%}  \\ 
$F_{(|0\rangle+i|1\rangle)/\sqrt{2}}$  & \SI{96.17\pm0.0005}{\%}  \\ 
$F_{(|0\rangle-i|1\rangle)/\sqrt{2}}$  & \SI{96.28\pm0.0079}{\%}  \\ 

\end{tabular}
\end{ruledtabular}
\end{table}

In quantum process tomography, the Choi matrix is reconstructed using the density matrices obtained from the quantum state tomography of the six cardinal states, as detailed in Section \ref{sec tomo}. The fidelities of the density matrices are presented in Table \ref{table cardinal fidelities}. The fidelities of states $F_{(|0\rangle+|1\rangle)/\sqrt{2}}$ and $F_{|1\rangle}$ presented in Table \ref{table cardinal fidelities} are comparatively lower than those in the last row of Table \ref{table fidelities}, potentially due to the increased probability of system drift when a larger dataset of six states is required and a constant global phase across states is needed.



\nocite{*}

\bibliography{apssamp}

\end{document}